\documentclass{elsart1p}
\usepackage{graphicx}
\usepackage{amssymb}
\usepackage{amsmath}
\begin{document}
\begin{frontmatter}
% notes:
% use the corauthref command within \author for corresponding author
% footnotes;
\title{Open charm measurement in p+p $\sqrt{s}$ = 200 GeV collisions at STAR}

\author{David Tlust\'y for STAR collaboration}
\address{FNSPE, Czech Technical University in Prague, Prague, Czech 
Republic}
\address{Nuclear Physics Institute AS CR v.v.i, Prague, Czech Republic}
\begin{abstract}
The charm production is sensitive to early dynamics of the created system
in RHIC heavy ion collisions. Dominant process of charm quarks production 
at RHIC is believed to be initial gluon fusion which can be calculated in the 
perturbative QCD. Understanding both the charm production total cross 
section and the fragmentation in p+p collisions is a baseline to further exploring 
the QCD medium via open charm and charmonium in heavy ion collisions.
In this paper we present the first reconstruction of open charm meson $D^{0}$
via the weak decay to K and $\pi$ mesons in the p+p collisions at $\sqrt{s}$ = 200 GeV.
The analysis is based on the large p+p minimum bias sample
collected in RHIC year 2009 by the STAR detector. 
The $D^0$ decay daughter identification was improved by using
the data of the newly installed Time-Of-Flight detector with
72\% of its designed coverage.

\end{abstract}
\vspace{-7mm}
\end{frontmatter}

% main text
\vspace{-7mm}
\section{Introduction}
\label{}

The charm quark production is dominated by initial gluon fusion at initial hard partonic collisions 
during heavy ion collisions. This production can be described by perturbative QCD (pQCD) due to
the large mass ($\sim$ 1.5 GeV/c$^2$) of the charm quark \cite{2}. Its mass is given by Electroweak
spontaneous symmetry breaking and the QCD chiral symmetry breaking does not affect charm quark mass
\cite{7}. Unlike light quarks (u,d,s), charm quarks cannot be easily produced during the mixed and hadronic phases of the dense matter
since the charm mass is much larger than the corresponding temperature scale \cite{4} at RHIC 
collision energies. In gluon radiative energy loss mechanism, charm quarks suffer less energy loss while
traversing through the dense partonic matter due to Dead Cone Effect \cite{6}. All these phenomena 
make charm quark an excellent probe of the Quark Gluon Plasma. 
Open charm analysis in p+p collisions is a crucial baseline to further explore 
the QCD medium via open charm and charmonium in heavy ion collisions.

\vspace{-1mm}
\section{The Measurement}
\begin{figure}[!h]
\begin{center}
\vspace{-1mm}
\includegraphics[width=0.7\textwidth]{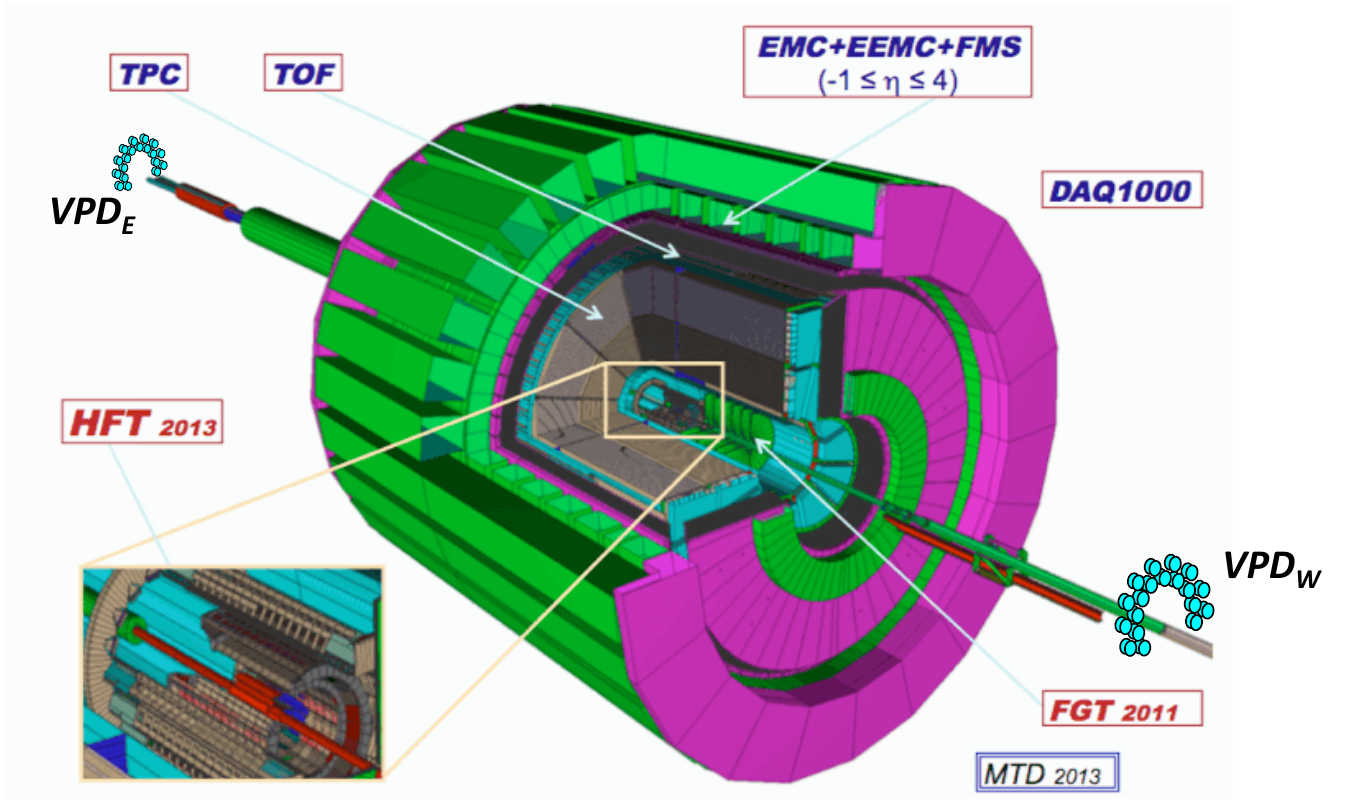} 
\end{center}
\vspace{-3mm}
\caption{The STAR detector. HFT and FGT are planned upgrades. TPC (Time Projection Chamber) is main detector
for tracking and PID (provides $\text{d}E/\text{d}x,\vec{p})$, TOF (Time Of Flight) is used for PID improvement
(time resolution 110 ps, start time is provided by VPD) and pileup tracks removal, BEMC (Barrel Electromagnetic
Calorimeter) used for pile-up tracks removal and high $p_\perp$ triggering.} 
\label{star}
\end{figure}

\vspace{-1mm}
Identification of charmed hadrons is difficult due to their short lifetime ($c\tau (D^0)$ = 123 $\mu$m), low 
production rates, and large combinatorial background. The open charm yield $\text{d}N_{D^0}^{pp}/\text{d}y$
is calculated from $D^0$ meson invariant mass reconstruction directly through the hadronic decay channel 
$D^0(\overline{D^0})\xrightarrow{\Gamma_i/\Gamma=0.038} K^\mp\pi^\pm$ and from $D^*$ through the invariant mass difference between $D^*$ and $D^0$ taken from 
 $D^{*\pm}\xrightarrow{\Gamma_i/\Gamma=0.677}D^0(\overline{D^0})\pi^\pm$
decay. In what follows, we imply $(D^0+\overline{D^0})/2$ when using the term $D^0$ unless otherwise specified.
The data used in this analysis were taken during the 2009 RHIC run in p+p collisions at  
$\sqrt{s}$ = 200 GeV with the Solenoidal Tracker at RHIC (STAR) (Fig. \ref{star}.).
These data are 107.7 million minimum bias 
data collected with the coincidence requirement on the east and west 
Vertex Position Detectors VPD, the VPDMB data.
 
The primary tracking device of the STAR detector is the Time Projection Chamber (TPC) \cite{9}. 
Candidate tracks were selected having momenta $p (p_\perp) > 0.3 (0.2)$ GeV/c and pseudorapidity $|\eta|<1$.
TPC was used to reconstruct the $D^0$ decay with the help of newly installed Time of Flight Detector (TOF) \cite{10}. TOF was used as a main PID tool to distinguish
kaons from pions (up to $p_\perp = 1.6$ GeV/$c$). 
Pions were identified 
via $\text{d}E/\text{d}x$ provided by TPC as kaon contamination is small and requiring 
TOF PID for pions given limited TOF acceptance reduces the expected 
signal significance.

\vspace{-1mm}
\subsection{$D^0$ raw yield}
\vspace{-3mm}

\begin{figure}[!h]
\begin{center}
\includegraphics[width=0.7\textwidth]{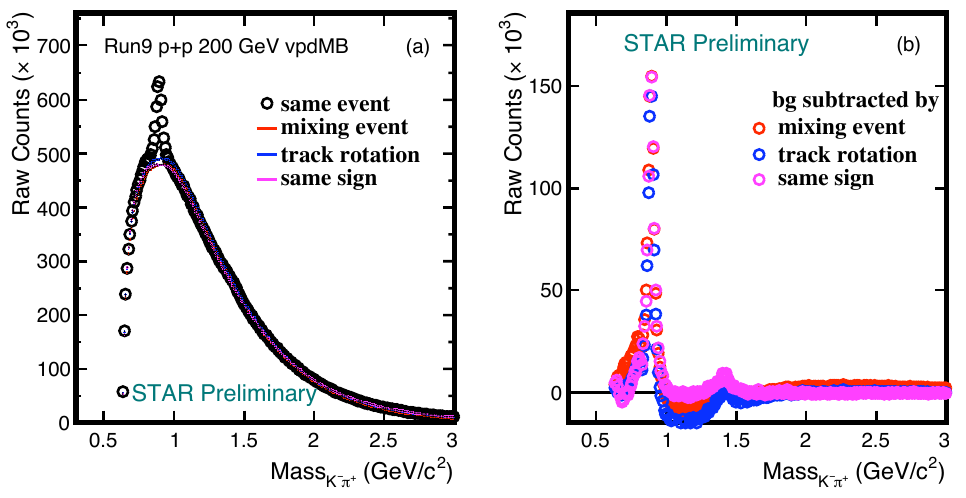} 
\end{center}
\vspace{-4mm}
\caption{(a) Invariant mass distributions of $K$-$\pi$ 
pairs from p+p collisions with combinatorial 
background reconstructed with 3 different methods. (b) Invariant mass after combinatorial background
subtraction
} 
\label{invmass}
\end{figure}

\begin{figure}[!h]
\begin{center}
\includegraphics[width=0.7\textwidth]{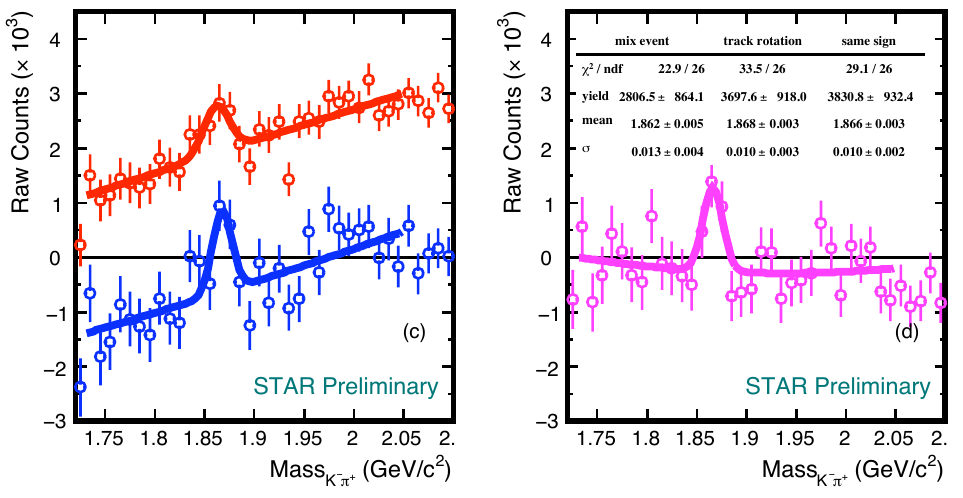} 
\end{center}
\vspace{-4mm}
\caption{(a) Invariant mass distributions of $K$-$\pi$ 
pairs from p+p collisions after combinatorial background
subtraction reconstructed through mixed event, track rotation (b)
and same sign method.} 
\label{zoom}
\end{figure}

From selected candidates, pairs were created and their
invariant mass was reconstructed. 
The invariant mass spectrum of $K\pi$ pairs are presented in Fig. \ref{invmass}. The combinatorial background
is reconstructed with three different methods:
\begin{enumerate}
\item \textbf{Mixed event}: Events are categorized according to the 
z-position of event vertices. Pions from one event are paired with kaons 
from other random events from an
event pool with similar global features. 
\item \textbf{Track rotation}: Each $\pi$ is paired with $K$ with reversed 3-momenta (within same event). 
\item \textbf{Same Sign}: pions are paired with same charged kaons (within same event). The geometric mean for positive $N_{++}$ and negative $N_{--}$
pair is calculated as $2\sqrt{N_{++}N_{--}}$.                
\end{enumerate}
There are resonances $K^{*0} (892)$ and $K_2^{*} (1430)$ clearly visible in Fig. \ref{invmass}. To see
a $D^0$ peak, the zoom to mass window from 1.72 to 2.1 GeV/c$^2$ is necessary, as seen in Fig. \ref{zoom}. 

As we see further from Fig. \ref{zoom},  mixed event method
isn't able to fully reconstruct the whole background (only the combinatorial one).  
Same sign and track rotation reconstruct even non-combinatorial background sources additionally. 
From track rotation and same sign methods we obtain 4$\sigma$ signal. 
The final results of open charm raw yield look quite consistent (within stat. 
errors) among these three methods,as seen in Table 1. 

\begin{table}[!h]
\begin{center}
\includegraphics[width=0.6\textwidth]{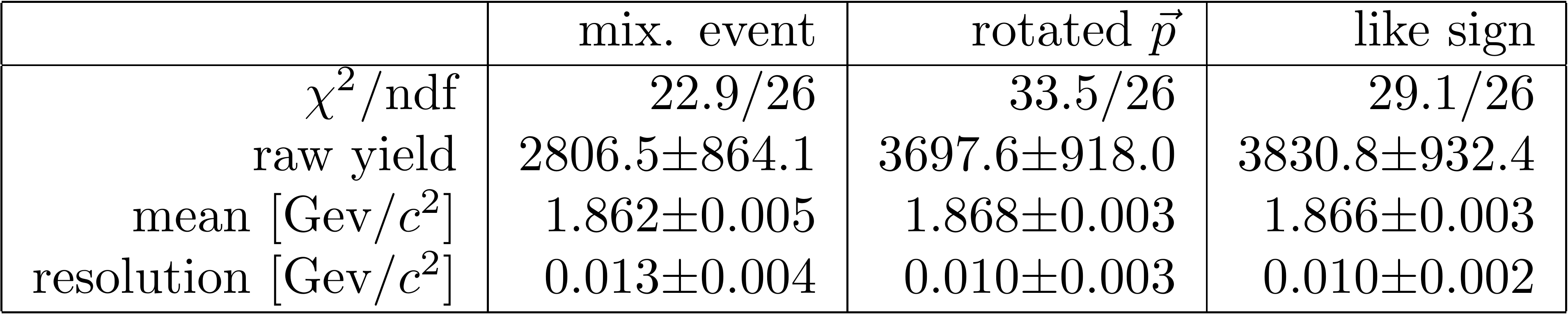} 
\end{center}
Table 1: Parameters of $D^0$ raw yield (after comb. bg. substraction) gausian fit.
The residual background is parametrized by second degree polynomial.
\end{table} 

\vspace{-3mm}
\subsection{$D^*$ raw yield}        
\vspace{-3mm}

\begin{figure}[!h]
\begin{center}
\vspace{-2mm}
\includegraphics[width=0.8\textwidth]{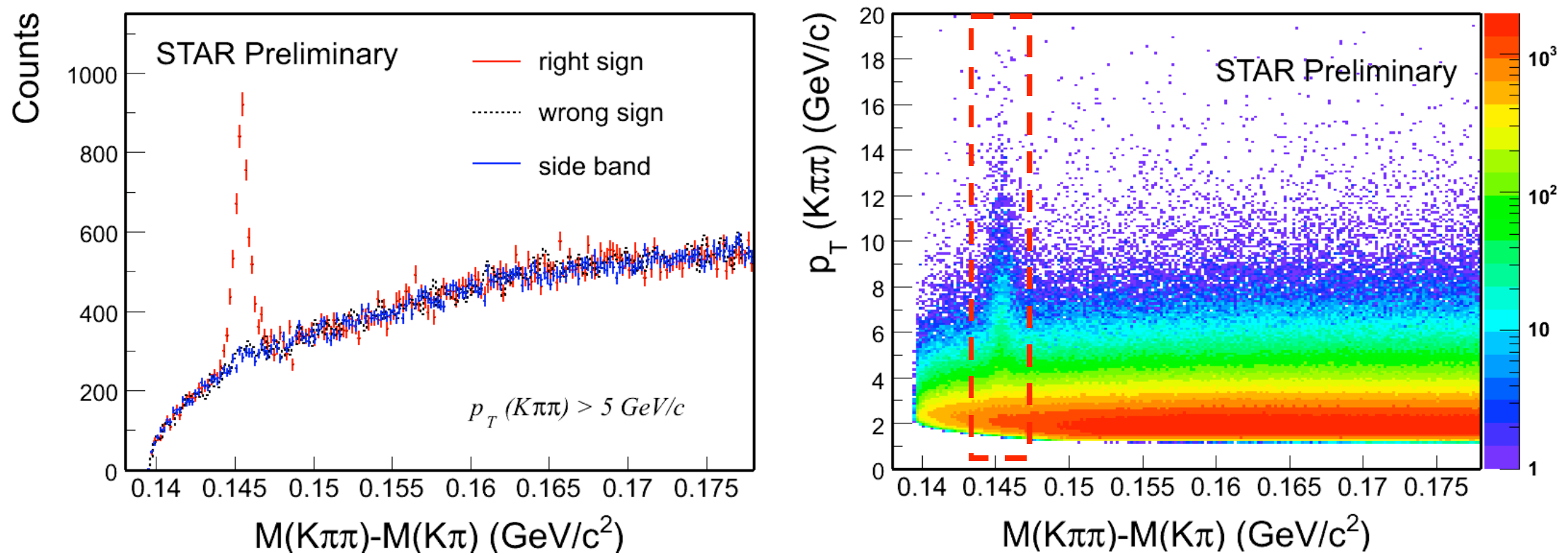} 
\end{center}
\vspace{-3mm}
\caption{(a) Invariant mass of $K\pi\pi - K\pi$  
differences from p+p collisions with combinatorial background
reconstructed through side band and wrong side techniques for high $p_\perp$ 
$D^*$ candidates (b) and scatter plot $p_\perp$ vs inv. mass for for all triggers.} 
\label{dstar}
\end{figure}

The $D^*$ decay has a small $Q$-value. The $D^0$ thus carries most of the $D^*$ momentum
and the $\pi$ from the $D^*$ decay, denoted by $\pi_s$, is soft. The enhancement in the distribution of 
the invariant mass difference $\Delta M = M(K^\mp\pi^\pm\pi^\pm_s)-M(K^\mp\pi^\pm)$ is used to
determine the $D^*$ raw yield \cite{11}.  A mass interval $1.82<M(K^\mp\pi^\pm)<1.9 \text{GeV}/c^2$ was
used to select $D^0$ candidates and two methods were used for combinatorial background  reconstruction, (see \cite{12} for details). Combinatorial
background was suppressed by requiring the $7<p_\perp(D^0)/p_\perp(\pi_s)<20$ and the decay angle of
the $K$ in the $K^\mp\pi^\pm$ rest frame, $\theta^*$, to be restricted by requiring $\cos(\theta^*)<0.8$
to remove near-collinear combinatorial background from jet fragmentation. 
The data sample used in the $D^*$ reconstruction includes the
minimum bias and the Jet-Patch triggered data. The
triggered data requires the sum of transverse energies deposited
in one patch of BEMC towers to be above a certain threshold. Result with all triggers is depicted in Fig. \ref{dstar}. 

\vspace{-4mm}
\section{Summary} 
\vspace{-3mm}

Newly installed TOF detector impoves particle identification in STAR significantly.
The 4$\sigma$ signal of $D^0$ meson in hadronic decays has been observed in p+p $\sqrt{s}=200$ GeV  collisions as well as strong  $D^*$ signal.
Further efficiency and acceptance corrections are being carried out to extract 
the charm hadron spectra as well as the total charm cross section. \footnote{This work was supported by grant INGO LA09013 of the Ministry of Education, Youth and Sports of the Czech Republic and by the Grant Agency of the Czech Technical University in Prague, grant No.SGS10/292/OHK4/3T/14}


\vspace{-5mm}
\begin{thebibliography}{99}
\vspace{-4mm}
\bibitem{2}{M. Cacciari, P. Nason and R. Vogt, \textit{Phys. Rev. Lett.} \textbf{95}, 122001 (2005).}
\bibitem{7}{X. Zhu \textit{et al., Phys. Lett.} B \textbf{647}, 366 (2007).}
\bibitem{4}{P. L\'evai, B. M\"uller and X. Wang, \textit{Phys. Rev.} C \textbf{51}, 6 (1995).}
\bibitem{6}{D. Kharzeev \textit{et al., Phys. Lett.}, B \textbf{519}, 199 (2001).}
\bibitem{9}{M. Anderson \textit{et al., Nucl. Instr. Meth.} A \textbf{499}, 659 (2003).}
\bibitem{10}{J. Wu \textit{et al., Nucl. Part. Phys.} \textbf{34}, S729 (2007).}
\bibitem{11}{S. Nussinov, \textit{Phys. Rev. Lett.} \textbf{35}, 1672 (1975).}
\bibitem{12}{B.I. Abelev \textit{et al., Phys. Rev.} D \textbf{79}, 112006 (2009).}


\end{thebibliography}
\end{document}